\documentclass[twocolumn,prl,showpacs,floatfix]{revtex4}

\usepackage{graphicx}
\usepackage{amsmath}

\begin{document}

\title
{
Segregation and phase inversion of strongly and weakly fluctuated Brownian
particle mixtures in spherical containers
}

\author
{
Akinori Awazu$^{1,2}$
}

\affiliation
{
$^1$Department of Mathematical and Life Sciences, Hiroshima University,
$^2$Research Center for Mathematics on Chromatin Live Dynamics. \\
Kagami-yama 1-3-1, Higashi-Hiroshima 739-8526, Japan.
}

\date{\today}

\begin{abstract}
We investigate the segregation pattern formations of strongly and weakly
fluctuated Brownian particle mixtures, which are confined in spherical
containers with finite volumes. We consider systems where the container
restricts the motions of particles combining two familiar methods: spherically
symmetric linear potential and the container edge wall. In such systems, the
following two segregation patterns are observed.  When the container radius
is large enough, more weakly fluctuated particles accumulate around the
center of the container than strongly fluctuated particles. On the other
hand, inverted distributions of strongly and weakly fluctuated particles are
observed when the container radius is small.  We also found a similar
segregation and phase inversion if such particle mixtures construct a
chain (hetero-fluctuated polymer) and are confined in a container with
no linear potential. We provide the qualitative mechanism and the
relationships for the biopolymer behaviors of such phase inversions.
\end{abstract}


\maketitle

Recently, the formation of spatially ordered patterns of self-propelled
populations such as bio-molecules, cells, animals, and artificial autonomous
motors have been extensively studied \cite{rev1,nagai0,nagai00,nagai,cell1,
cell2,cell3,cell4,cell5,awa,animal1,animal2,animal3,animal4,midori,nakata1,
nakata2,nakata3,nakata4,active1,active2,ganai}. Theoretical studies using
idealized models have provided several insights for the understanding of
physical, biological, and social phenomena.

Segregation patterns of particles involving different characteristics such as
size, shape, stiffness, and mobility are typical examples of pattern formations
that have been universally observed in nonequilibrium systems \cite{cell1,
cell2,cell3,active1,active2,ganai,kona1,kona2,kona3,kona4,kona5}. In most of
these studies, the primary focus is on the contributions of inhomogeneous
particles in bulk systems. On the other hand, in real systems, several objects
are usually confined in a limited space. For example, bio-molecules in cells
are often confined in subcellular organelles, and the structural stability
and reaction activities of bio-molecules are much different between in vivo
and in vitro situations\cite{ganai,crowd1,crowd2,crowd3,Aoki,Fujii}. During
the developmental processes of multi-cellular organisms, each cell migrates
in the restricted space in their host body\cite{hassei}. Moreover, the
segregation patterns of granular mixtures are also highly influenced by the
form and the size of the container\cite{kona2,kona4}. Thus, to uncover
pattern formations mechanisms in real nonequilibrium systems, the influence
of container characteristics should also be clarified.

In this letter, through simple nonequilibrium particle models, we investigate
the influence of the container characteristics, in particular container
size, on the segregation patterns. First, we consider the segregations of
\emph{hetero-Brownian particles} model consisting of finite volume particles
driven by different magnitude fluctuations. This model is considered to be one of
the simplest descriptions of active and inactive self-propelled particle
mixtures. We focus on the segregation pattern of strongly and weakly
fluctuated particles under container restrictions given by the combination
of two familiar effects: spherically symmetric linear potential in the bulk
and a wall at the edge of the container.

In addition, we consider the \emph{hetero-fluctuated polymer}, which is a
chain constructed by hetero-Brownian particles in a spherical container
without a linear potential. Such polymers are regarded to be a simplified
chromosome model for nuclei involving transcriptional active and inactive
(silenced) regions\cite{ganai}. Here, the regions containing strongly
fluctuated particles are considered to be transcriptional active regions
because several proteins such as chromatin remodeling factors and
transcription factors often access such DNA regions and produce several
mechanical perturbations through the ATP hydrolysis energy consumption.
Then, by focusing on the container size dependent behaviors of this model,
we understand the possible contributions of bio-molecule confinement in
biological activities.

We now introduce the model for hetero-Brownian particles and a
hetero-fluctuated polymer in a container. These systems consist of $N$
spherical particles with the common diameters of $d$. Particles are driven
by random forces, which have different average magnitudes for each respective
particle. Then, the equation of the motion of each particle is given by

\begin{equation}
{\bf \dot{x}_i} = -\nabla_i (V_{int}(\{ {\bf x}_i \}) + V_{con}(\{{\bf x}_i\})) + {\bf \eta}_i(t),
\end{equation}
\begin{equation}
<{\bf \eta}_i(t){\bf \eta}_i(t')> = 2G_{i}\delta(t-t'),
\end{equation}
where ${\bf x}_i$ is the position of the $i$th particle, ${\bf \eta}_i $ and
$G_i$ are the random force working on the $i$th particle and its magnitude,
respectively. Here, the origin (${\bf x}_i =0$) is the center of
the container.

The interaction potential between particles is indicated by
$V_{int}(\{ {\bf x}_i \})$.
Due to the excluded volumes of particles in the hetero-Brownian particle
model, we only consider the soft-core repulsive potential as
$V_{int}(\{ {\bf x}_i \}) = V^{sf}(\{ {\bf x}_i \})$, where
\begin{equation}
V^{sf}(\{ {\bf x}_i \}) = \sum_{i < j}
\begin{cases}
  \frac{k_e}{2}(|{\bf x}_i-{\bf x}_j|-d)^2 \,\,\,\,\,\, (|{\bf x}_i-{\bf x}_j| < d) \\
  0 \,\,\,\,\,\, (Otherwise)
\end{cases},
\end{equation}
with the elastic constant $k_e$. In the hetero-fluctuated polymer model, we
assume that chain of Brownian particles does not have branches. Then, we have
$V_{int}(\{ {\bf x}_i \})=V^{sf}(\{ {\bf x}_i \}) + V^{ch}(\{ {\bf x}_i \})$,
where
\begin{equation}
V^{ch}(\{ {\bf x}_i \}) =  \sum_{i}\frac{k_c}{2}(|{\bf x}_i-{\bf x}_{i+1}|-d)^2
\end{equation}
with constants $k_e$ and $k_c$.

The potential for the container spatial constraints is given by
$\displaystyle V_{con}(\{{\bf x}_i\})$. In this letter, we consider
$\displaystyle V_{con}(\{{\bf x}_i\}) = \sum_i V^i_{con}({\bf x}_i)$,
where $V^i_{con}$ is given by sum of the potential of the wall at the edge
of container $V^i_{wall}$ and the linear potential in the bulk $V^i_{bulk}$
(see Fig. 1(a)) as
\begin{equation}
V^i_{wall}({\bf x}_i) =
\begin{cases}
  \frac{k_w}{2}(|{\bf x}_i|-(R-\frac{d}{2}))^2 \,\,\,\,\,\, (|{\bf x}_i| > R-\frac{d}{2}) \\
  0 \,\,\,\,\,\, (Otherwise)
\end{cases}
\end{equation}
and
\begin{equation}
V^i_{bulk}({\bf x}_i) = k_l|{\bf x}_i|
\end{equation}
with constants $k_w$, $k_l$, and container radius $R$.

We now focus on the simplified systems containing only two types of particles:
strongly fluctuated particles (S-particles) with $G_i = G_s$ and weakly
fluctuate particles (W-particles) with $G_i = G_w$. Here, we give the number
of S- and W-particles as $N/2$,  $G_s = 1$,  $G_w = 0$, $k_e = k_c = 1024$,
$d = 1$, and $k_l$ holding $G_s / k_l >> d^2 $. Even when $G_w > 0$, we
qualitatively obtain the same results as the case when $G_w << G_s$ holds.
In a hetero-fluctuated polymer, we also assume that the S- and W-particles
are periodically connected where the length of each S- and W-particles region
is $L/2$ (see Fig. 1).

\begin{figure}
\begin{center}
\includegraphics[width=7.0cm]{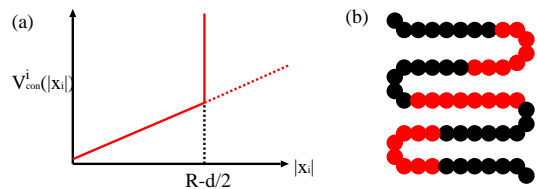}
\end{center}
\caption{(Color online) (a) Illustration of the container potential.
(b) Illustration of hetero-fluctuated polymers where the black and gray (red)
circles indicate weakly and strongly fluctuated particles, respectively.}
\end{figure}

\begin{figure}
\begin{center}
\includegraphics[width=7.0cm]{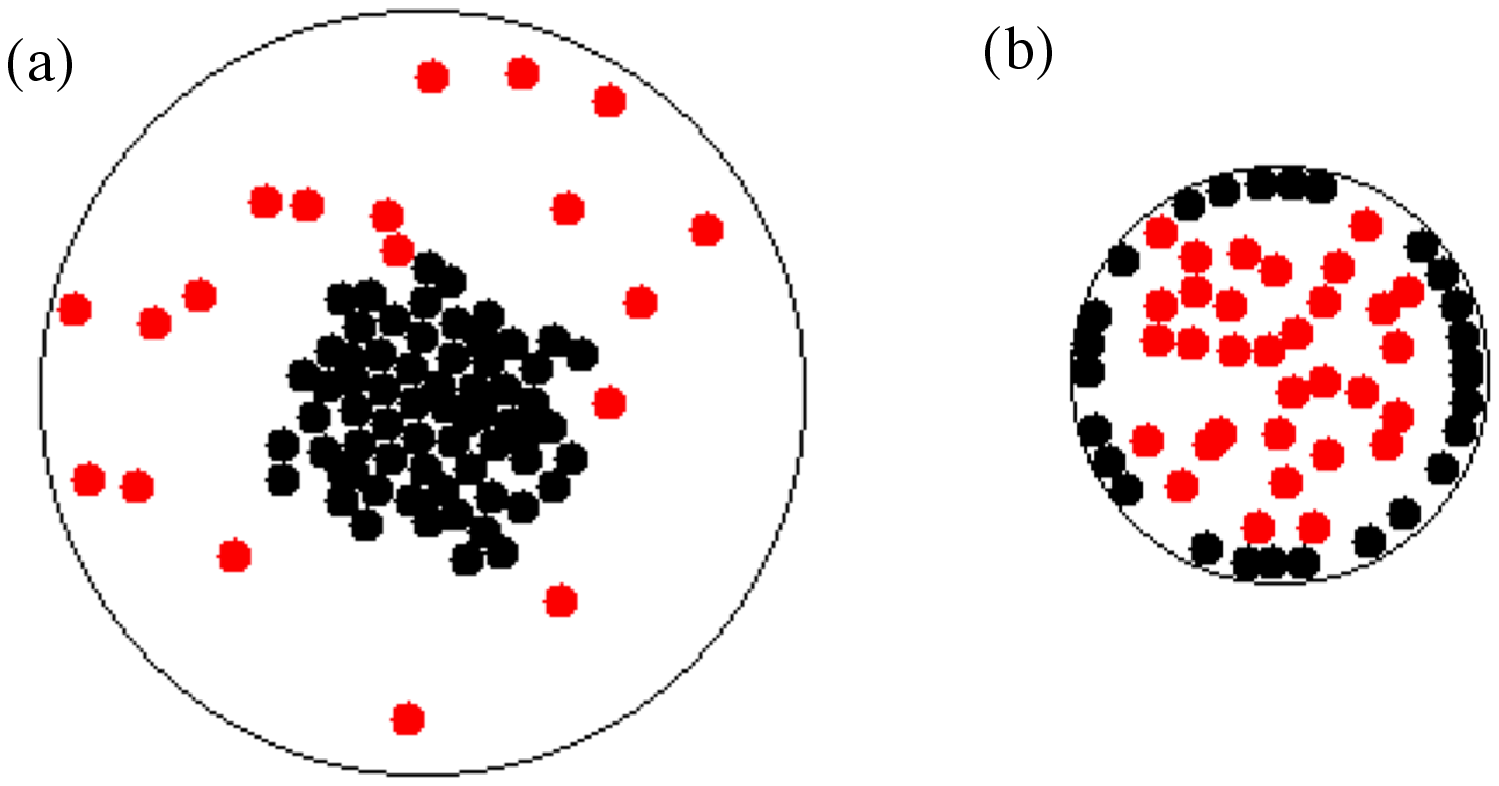}
\includegraphics[width=5.0cm]{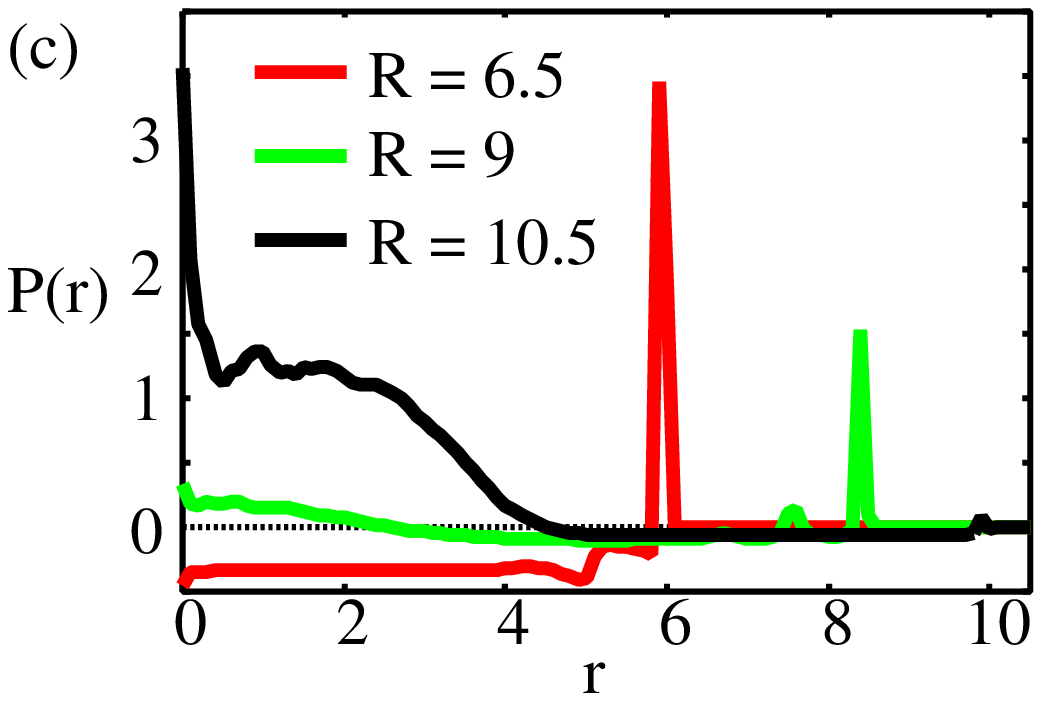}
\end{center}
\caption{(Color online) Typical snapshots of the distributions of S-particles
(grey (red)) and W-particles (black) of hetero-Brownian particles model
(N=512) on the 2-D cross section for (a) $R=10$, (b) $R=5$, and
(c) $P(r)$ for typical $R$.}
\end{figure}

First, we focus on the segregation pattern formations of hetero-Brownian
particles in the container. In the case of $k_l > 0$ and $k_w=0$,
W-particles (S-particles) tend to locate near (far from) the container center,
which may be easily expected by considering the $G_i$-dependent distribution
of particles affected by the potential. On the other hand, if there exists
a hard wall at the edge of the container, the steady state properties of the
system changes as follows.

Figures 2(a) and 2(b) show typical snapshots of the S- and W-particle
distributions in the 2-D cross section (particles at $-d/2 \le x \le d/2$
on the $x-y-z$ space are shown) for $R=11$ and $R=6$, respectively. Figure 2(c)
shows the relative radial distributions $P(r)$ for some $R$ with $k_l = 0.01$,
$k_w = 1024$, and $N=512$. Here, $P(r)$ is given as $P_w(r)-P_s(r)$, where
$P_m(r) = n_m(r)/4\pi r^2$ ($m =$ S or W) are the respective radial particle
distributions, $r$ is the distance from the origin, and $n_m(r)$ is the
frequency of $m$-particles in the region between $r$ and $r +dr$
($dr = 0.1$ is employed). As shown in these figures, the distributions of
S- and W-particles are highly influenced by $R$: more W-particles are
distributed around the center than S-particles for large enough $R$, while
the inverted distribution occurs for small $R$.

The phase inversions are independent of $k_l$, $D_s$, and $N$ even though
the $R$ value at which phase inversion occurs ($R^*$) depends on these values.
To understand these dependencies, we study the simplest possible system with
$N=2$, which contains one S-particle and one W-particle. Figure 3(a) shows
$n(r) = n_w(r)-n_s(r)$ \cite{shuusoku} for several $R$ with $k_l=0.01$, while Fig. 3(b) shows
$R^*$ as a function of $k_l$. Here, $R^*$ is given as the maximum $R$ with
the peak of $n(r)$ near the edge of container. As shown in these figures,
$R^*$ is given by $\sim k_l^{-1/5}$.

To consider the mechanism of such results, we first examine the case $k_l=0$.
In this case, the motions of W-particles are driven only by collisions with
S-particles. Then, once W-particles reach the edge wall, the S-particle
forces acts only in the direction from the center to the outer wall. Thus,
the W-particles tend to stay at the edge wall independent of $R$.

Based on the above fact, $R^*$ is roughly estimated as follows. For simplicity,
we assume $D_s$ is so large compared to $k_lR$ that the influence of the
linear potential is negligible for S-particles. Basically, the W-particles
tend to move to the container center with velocity $k_l$ by the influence
of the linear potential. However, the W-particles tend to move to the edge
of the container if the S- and W-particles collide frequently. The collision
rate between S- and W-particles is proportional to two values: the volume
fraction of particles (order $(d/R)^{-3}$) and the inverse of the time that
the S-particles entire the space (order $G_sR^{-2}$). Here, we assume that
the average distance of the W-particle's outward-directed movements during 
a single collision is $\lambda$. Then, W-particles tend to move 
to the container edge if the collision rate is so large that $d^3G_sR^{-5}
\lambda > k_l$, which indicates $R^* \sim (d^3G_s \lambda/k_l)^{1/5}$.

We obtained similar results if $V^i_{bulk}$ is a harmonic potential.
However, if $V^i_{bulk}$ is given by $ ~ (|{\bf x_i}|/R_o)^n$ with a large
$n$, more S-particles tend to occupy the interior positions of the container
than W-particles independent of $R$ and $R_o$ because $V^i_{bulk}$ is regarded as the
potential indicating a wall exists at $|{\bf x_i}|=R_o+d/2$.

\begin{figure}
\begin{center}
\includegraphics[width=8.0cm]{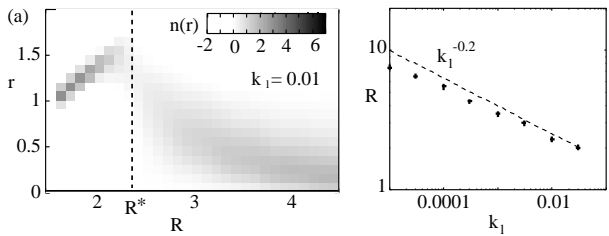}
\end{center}
\caption{(a) $n(r)$ for several $R$ with $k_l=0.01$, and (b) $R^*$ as a
function of $k_l$ for $N=2$ with dashed curve indicating $R^* = k_l^{-1/5}$.}
\end{figure}

\begin{figure}
\begin{center}
\includegraphics[width=7.0cm]{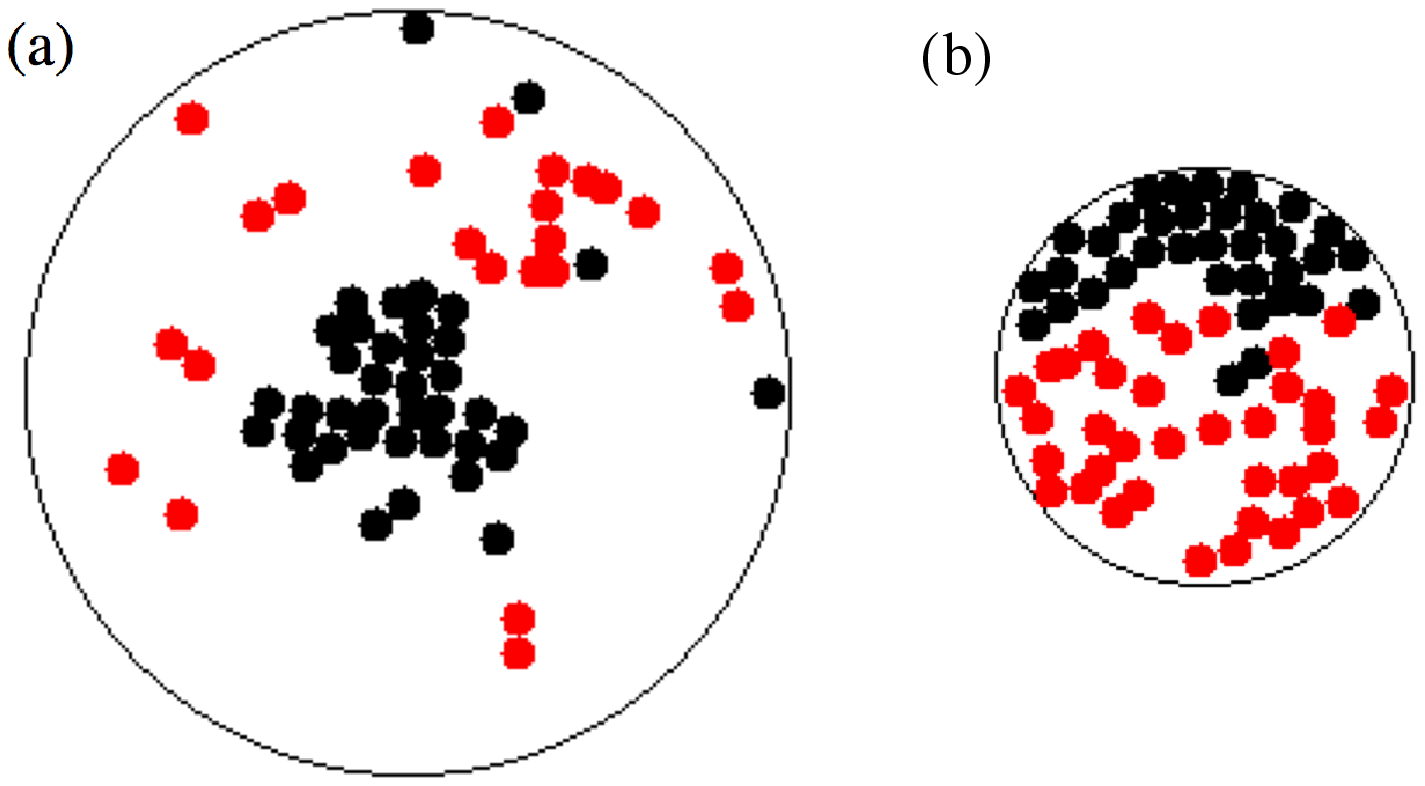}
\includegraphics[width=8.0cm]{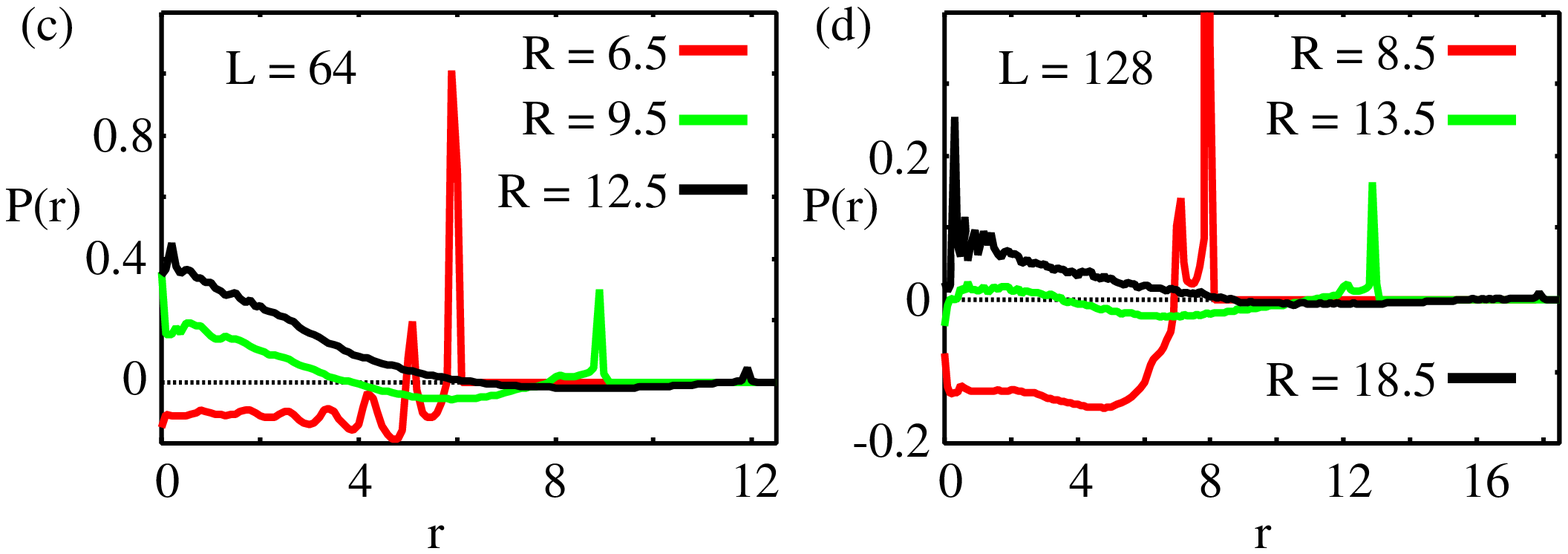}
\end{center}
\caption{(Color online) Typical snapshots of the distributions of
S-particles (grey (red)) and W-particles (black) of the hetero-fluctuated
polymer model (N=512) on the 2-D cross section for (a) $R=11$
and (b) $R=6$ with $L=64$, and $P(r)$ of this model for typical $R$ 
with (c) $L=64$ and (d) $L=128$.}
\end{figure}

Next, we focus on the segregations of S- and W-particles constructing the
hetero-fluctuated polymer in the spherical box ($k_l = 0$ and $k_w = 1024$).
In the following, we focus on the cases of $L = 32, 64, 128$ for $N = 512$, respectively. 
Figures 4(a) and 4(b) show typical snapshots of the
particle distributions on the 2-D cross-section with $L=64$ for
$R = 11$ and $R = 6$, respectively, and Fig. 4(c) and 4(d) show the relative radial
distribution functions of steady state S- and W-particles for some $R$ with $L=64$ 
and $L=128$, respectively. 
As shown in these figures, more W-particles tend to be distributed at the container 
center than S-particles for large enough $R$. On the other hand, W-particles tend 
to be distributed near the edge of the container for small $R$.

In a recent study, the latter type of segregation pattern was observed as
"activity-based segregation" in a similar model \cite{ganai}. On the other
hand, the present result indicates the patterns induced by the segregations
generally depends on the container size. We found that phase inversion occurs
at $R = 9 \sim 10$ in the case of $L=64$ and at $R = 13 \sim 14$ in the
case of $L = 128$. We also found that the latter segregation pattern does not 
appear when $L=32$. Then, these results indicate phase inversion occurs at 
$R \sim (dL/2)^{3/5}+d$ only if $L$ is large enough. We found almost the same 
results in the cases of $N = 256$ and $1024$. 

In order to qualitatively explain the present phase inversions, we propose
the following scenario. In general, the Brownian motion of a chain induces
the entropic force (rubber elasticity), which works on each chain element
to promote chain assembly.

In an equilibrium system, the elasticity of effective force is proportional
to the temperature. However, the hetero-fluctuated polymer consists of
particles with large and small fluctuations with $G_s$ and $G_w$,
respectively, that is equivalent to chains containing parts with high and
low temperatures. Then, the effective elasticity for each particle seems to
be influenced by both $G_s$ and $G_w$. Here, it seems natural that the
magnitude of the elasticity for $i$th particle is proportional to $G_i^*$,
which depends on $i$ but always holds for $G_w < G_i^* < G_s$.

Then, if we can neglect the effects of the excluded particle volume and the 
edge wall of the container, the radial distribution of the S- and W-particles is 
approximately obtained by a 3-D Gaussian distribution around the polymer center of the mass 
with the variance proportional to $G_s / G_i^*$ and $G_w / G_i^*$, respectively.
Even when each particle has the finite volume, the diversity of the S- and
W-particles distributions highly correlates with $G_s / G_i^*$ and
$G_w / G_i^*$, respectively. Thus, if $R$ is large enough, more W-particles
are located near the center of the polymer than S-particles since
$G_w / G_i^* < 1 < G_s / G_i^* $ holds for all $i$.

The above arguments indicate that W-particles (S-particles) tend to be
distributed near (far from) the container center because the center of mass
of the polymer tends to close to the container center. 
Moreover, when $R$ is sufficiently large, the distance between a particle at 
the center of S-particle region and that of the neighboring W-particle regions 
is estimated $\sim (dL/2)^{3/5}$ on average by the arguments of self-avoiding 
random walk \cite{dojan}. 
Then, when $R < (dL/2)^{3/5}$, the S- and neighboring W-particle regions tend to
collide excessively with each other while the force assembling the regions
becomes weaker. In such cases, the W-particles tend to move to
the edge of the container by similar mechanisms with those obtained in the
hetero-Brownian particle system.

If the volume fraction of the polymer is close to that of the closed packing, 
some corrections are needed to this arguments. However, the volume fractions 
of our simulations ($N = 256, 512$ and $1024$) are considered small enough when 
$R \sim (dL/2)^{3/5}$. Thus, the phase inversion $R$ seems insensitive 
for $N$ in our simulations.

In this letter, we investigated the segregation patterns of strongly and
weakly fluctuated Brownian particles confined in a spherical container. We
found that the segregation patterns of such systems drastically depend
on the container. Thus, in order to uncover the universal and individual
aspects of several nonequilibrium particle populations, we should consideration 
of the effects of several containers.

The size of the cell nucleus depends on the type and the developmental stage
of the cell, even though the chromosome volume in each cell is the same. Then,
the results of the presented polymer model provide some insights into cell
type-dependent chromosome behaviors. On the other hand, for pattern formations
of a confined chain, we also note that the elasticity and the heterogeneity
of the chain play important roles\cite{hermann,pcook}. Then, for more
detailed arguments for the intra-nucleus chromosome structures, we will
consider the hetero-fluctuated polymer model with further modifications by
referencing recent studies.

The author is grateful to T. Sugawara, S. Shinkai, S. Lee, T. Sakaue and
H. Nishimori for fruitful discussions and useful information. This research
was supported in part by the Platform for Dynamic Approaches to Living System
from the Ministry of Education, Culture, Sports, Science and Technology,
Japan, and the Grant-in-Aid for Scientific Research on Innovative Areas
(Spying minority in biological phenomena (No.3306) (24115515)) of MEXT of Japan.


\begin{thebibliography}{9}
\bibitem{shuusoku}
Here, $n(r)$ is employed instead of $P(r)$ because the convergence of $P(r)$ is slow and numerically unstable for small $r$ in cases of small $N$.
\end{thebibliography}
\end{document}